\documentclass[12pt]{article}
\usepackage{amsmath, amssymb, amsthm, dsfont}
\usepackage{booktabs} 
\usepackage[table]{xcolor} 
\usepackage{mathrsfs}
\usepackage{hyperref}
\usepackage{mathscinet}
\usepackage{setspace} 
\usepackage[indent]{parskip}

\usepackage[
  height=22cm,
  width=15.5cm,
  top=2.5cm,
  hmarginratio=1:1
]{geometry}

\numberwithin{equation}{section}
\theoremstyle{definition}
\newtheorem{thm}{Theorem}[section]

\newcommand{\cE}{\mathcal{E}}
\newcommand{\cN}{\mathcal{N}}
\newcommand{\cO}{\mathcal{O}}

\newcommand{\SL}{\mathrm{SL}}

\newcommand{\SU}{\mathrm{SU}}
\newcommand{\Sp}{\mathrm{Sp}}
\newcommand{\SO}{\mathrm{SO}}
\newcommand{\Spin}{\mathrm{Spin}}

\newcommand{\R}{\mathbb{R}}
\newcommand{\bC}{\mathbb{C}}
\newcommand{\bZ}{\mathbb{Z}}

\newcommand{\fl}{\mathfrak{l}}
\newcommand{\fz}{\mathfrak{z}}
\newcommand{\fsu}{\mathfrak{su}}
\newcommand{\fso}{\mathfrak{so}}
\newcommand{\fsp}{\mathfrak{sp}}

\DeclareMathOperator{\SYM}{SYM^{D=2}_{\mathcal{N} = (8,8)}}
\DeclareMathOperator{\CFT}{CFT}
\DeclareMathOperator{\elliptic}{ell}
\DeclareMathOperator{\Irr}{Irr}

\def\LG{{^L\negthinspace G}}

\newcommand{\lisom}{\stackrel{\sim}{\longrightarrow}}

\begin{document}
\begin{titlepage}

\vskip 2cm

\begin{center}
{\large \bfseries
Vacuum Structure of Maximally Supersymmetric Two-Dimensional Gauge Theories
}

\vskip 1.2cm
Richard Eager\footnote{rdeager@umd.edu}
\bigskip
\bigskip

Department of Physics \\
University of Maryland, College Park, MD 20742, USA

\vskip 1.5cm

\textbf{Abstract}
\end{center}
We study the vacuum structure of maximally supersymmetric Yang--Mills theory in two dimensions.  At low energies, the theory decomposes into superselection sectors consisting of orbifold conformal field theories and massive vacua.  The sectors are indexed by the cuspidal data appearing in Lusztig's generalized Springer correspondence.  This suggests an enhancement of the correspondence to holomorphic factorization algebras.
\medskip
\noindent

\bigskip
\vfill
\end{titlepage}


\newpage 
\section{Introduction}
Supersymmetric gauge theories have rich quantum dynamics.  Surprisingly, seemingly complicated gauge theories can be equivalently described in terms of simpler theories through dualities.  Montonen and Olive conjectured a duality
that interchanges the electric and magnetic charges between a gauge theory with gauge group $G$ and a dual theory with Goddard--Nuyts--Olive (GNO) dual gauge group $\LG$ \cite{Montonen:1977sn, Goddard:1976qe}.  Shortly thereafter, the conjecture was sharpened to an exact duality conjecture for four-dimensional $\mathcal{N} = 4$ supersymmetric Yang--Mills theory \cite{Osborn:1979tq}.
 
Sen argued that electric-magnetic duality requires the existence of supersymmetric bound states of monopoles and dyons \cite{Sen:1994yi}.  Similar phenomena in two-dimensional gauge theories can be realized directly in string theory.  Witten argued that the $SL(2, \bZ)$ invariance of type IIB string theory implies the existence of bound states of fundamental strings and D-strings \cite{Witten:1995im}.  This leads to the rather subtle prediction of threshold bound states in the world-volume theory of coincident D-strings.  In particular, string theory predicts the existence of a normalizable bound state in two-dimensional $\mathcal{N} = (8,8)$ supersymmetric Yang--Mills theory with special unitary gauge group.  

General arguments suggest that two-dimensional $\mathcal{N} = (8,8)$ supersymmetric Yang--Mills theory decomposes into superselection sectors of orbifold conformal field theories and massive vacua \cite{Seiberg:1997ax, Kologlu:2016aev}.  However, these arguments give no hint as to which orbifolds appear.  We propose that
for a simply connected gauge group $G$, the low-energy dynamics are described by the superselection sectors 
\begin{equation*}
\SYM(G) \rightarrow \coprod_{(L, \mathcal{E})} \CFT  \bigl( \left( \fz(\fl) \otimes \R^8 \right)/W_{G,L} \bigr).
\end{equation*}
Similarly, the Coulomb branch $\mathcal{M}(G)$ decomposes into a disjoint union
\begin{equation*}
\mathcal{M}(G) \rightarrow \coprod_{(L, \cE)} \left( \fz(\fl) \otimes \R^8 \right)/W_{G,L}.
\end{equation*}
In both cases, the disjoint union is over cuspidal data $(L, \mathcal{O}_L, \cE_L)$ consisting of a Levi subgroup $L$, a unipotent orbit $\cO_L$, and a cuspidal local system $\cE_L$ on $\cO_L$.
Each triple is uniquely determined by the pair $(\cO_L, \cE_L)$ up to simultaneous conjugacy by $G$. We denote the relative Weyl group by $W_{G,L}$ and the center of the Lie algebra of $L$ by $\fz(\fl)$.

This conjectural decomposition is inspired by Lusztig's generalized Springer theory \cite{MR732546}.  Let
$\mathcal{N}(G)$ denote the set of pairs $(\cO, \cE)$ with $\cO$ a unipotent conjugacy class of $G$ and $\cE$ an irreducible $G$-equivariant local system on $\cO$ up to isomorphism and denote the irreducible representations of the relative Weyl group by $\Irr W_{G,L}$.
\begin{thm}[Lusztig, Thm 6.5 \cite{MR732546}]
There is a bijection
\begin{equation*}
\mathcal{N}(G) \lisom \coprod_{(L, \mathcal{E})} \Irr W_{G,L}.
\end{equation*}
\end{thm}
Alternatively, we can view the low-energy equivalence as a further generalization of Lusztig's generalized Springer correspondence.  In the strongest form, it suggests an equivalence of holomorphic factorization algebras \cite{MR3586504, MR4300181, MR4971881}.  
For special unitary gauge groups, we recover recover
Kolo\u{g}lu's \cite{Kologlu:2016aev} decomposition
\begin{equation*}
\SYM(SU(N)) \rightarrow \coprod_{k=1,\dots, N} \CFT  \bigl( \R^{8(d-1)} /W_{S_d} \bigr), \qquad d = \gcd(k, N),
\end{equation*}
where $d=1$ contributes a single massive vacuum.
The sectors describe the dynamics of bound states of $(k/d, N/d)$ strings in Type IIB string theory.
Further evidence for the proposed low-energy dynamics comes from matching the bound states between the ultraviolet (UV) and infrared (IR), counting bound states in their M-theory and IIB orientifold realizations, and supersymmetric localization.

Supersymmetric localization has allowed for precision tests of string theory predictions about the quantum dynamics of two-dimensional gauge theories.  Its development was spurred in part by bound-state counting problems.
Our proposed equivalence implies the equality of the elliptic genus of the UV and IR theories \cite{Witten:1986bf, MR3177993, MR3302634, Gadde:2013dda}.  For special unitary groups, Kolo\u{g}lu computed the elliptic genus of both the UV and IR theories \cite{Kologlu:2016aev}.  The conjectural identity involves sums over four-dimensional partitions.  The elliptic genus receives contributions from both the bound states and continuum states.  A special limit of the elliptic genus recovers the bound states.

One fruitful approach to study the dynamics of supersymmetric gauge theories is to relate them to supersymmetric quantum mechanics.  For two-dimensional gauge theories, the bound states can be expressed as a sum over one-dimensional bound states.  From a modern perspective, the relationship between the two bound-state counting problems is governed by holonomy saddles \cite{Hwang:2017nop, Hwang:2018riu}.

Witten argued that M-theory required the existence of one supersymmetric bound state in $\mathcal{N} = 16$ supersymmetric quantum mechanics on the D-particle worldvolume \cite{Witten:1995im, Sen:1995vr}.  The prediction of threshold bound states in this theory was studied directly in supersymmetric quantum mechanics using a variety of methods including heat-kernel regularization, mass-deformations, and cohomological field theories \cite{Yi:1997eg, Sethi:1997pa, Porrati:1997ej, Green:1997tn, Moore:1998et, Kac:1999av}.  

In Section~\ref{sec:threshold}, we compute the number of bound states in the two-dimensional theory using holonomy saddles.  We then compute the number of threshold bound states in the IR orbifold description.  The equality of the number of threshold bound states of the IR and UV descriptions is the first non-trivial test of our proposal.  For classical gauge groups, the equalities follow from classical identities about partitions; for exceptional gauge groups, they are verified by direct computation.
  
In Section~\ref{sec:mtheory}, we sketch some results related to a possible M-theory interpretation.  The proposed orientifold description also gives a complete prediction for the refined index of $\mathcal{N} = 16$ supersymmetric quantum mechanics matching low-rank calculations of \cite{Lee:2016dbm} and the general unrefined result of \cite{Kac:1999av}.  The proposed IR description is also compatible with decomposition \cite{Hellerman:2006zs, Sharpe:2022ene}.

We conclude by reviewing some of the ingredients that are needed to check the equivalence of the UV and IR elliptic genera. We speculate on connections to elliptic Springer theory and an elliptic generalization of the Local Langlands Conjectures in Section~\ref{sec:elliptic}.

\section{Threshold Bound States}
\label{sec:threshold}
A bound state is a state that is normalizable except for the center of mass motion; it is at threshold if there is no energy barrier to its dissolution.
We first review the counting of threshold bound states in two-dimensional $\cN=(8,8)$ supersymmetric Yang--Mills theory on a circle.  We then count bound states in the proposed low-energy description of a disjoint union of orbifold conformal field theories.  Crucially, we will see that the two computations agree.
Many features of the bound-state counting problem are similar across dimensions, so we will consider $\cN = 16$ supersymmetric quantum mechanics, two-dimensional $\cN = (8,8)$ maximally supersymmetric Yang--Mills theory, and three-dimensional $\cN = 8$ supersymmetric Yang--Mills theory in parallel.  All of these theories can be obtained from dimensional reduction of 10-dimensional supersymmetric Yang--Mills theory \cite{Brink:1976bc}.  These bound state counting problems are simpler versions of bound state counting in four-dimensional $\cN = 4$ supersymmetric Yang--Mills theory.
  
Henningson and Wyllard determined the number of threshold bound states of $\cN=4$ supersymmetric Yang--Mills theory on $T^3$ in a series of papers \cite{Henningson:2007dq, Henningson:2007qr, Henningson:2008ri}.  We can conveniently and suggestively summarize the analogous results in dimensions one, two, and three using the invariants $n_d(G)$ defined by Jakob and Yun \cite{jakob2024countingabsolutelyindecomposablegbundles}.
The number of bound states of $\mathcal{N} = 16$ supersymmetric quantum mechanics, 2d $\mathcal{N} = (8,8)$ supersymmetric Yang--Mills theory on $S^1$, and 3d $\mathcal{N} = 8$ supersymmetric Yang--Mills theory on $T^2$ equals the invariant $n_d(G).$

For supersymmetric quantum mecahnics, $N_0(G)$ is the set of distinguished nilpotent orbits in $\mathfrak{g}$ and $n_0(\mathfrak{g})$ its cardinality, recovering Kac--Smilga's result \cite{Kac:1999av}.  For special unitary groups $n_0(\fsu_n) = 1$. For special orthogonal groups, $n_0(\fso_n)$ equals then number of partitions of $n$ into distinct odd parts.  For symplectic groups, $n_0(\fsp_{2n})$ equals the number of partitions of $n$ into distinct parts. 

The most relevant case to our story is the set $N_1(G)$ of strongly isolated conjugacy classes in $G$.
The cardinality $n_1(G)$ is the number of threshold bound states in $\mathcal{N} = 16$ supersymmetric quantum mechanics with gauge group $G.$
Let $\Sigma(G)$ be the set of isolated
semisimple conjugacy classes and $G^{\circ}_s$ be the connected centralizer of an element $s \in G$. Then $n_1(G)$ can be computed in terms of the $1d$ invariants through
\begin{align*}
N_1(G) & \leftrightarrow \coprod_{[s] \in \Sigma(G)} N_0(G^{\circ}_s) \\
n_1(G) & = \sum_{[s] \in \Sigma(G)} n_0(G^{\circ}_s). \\
\end{align*}
For special unitary gauge groups, we have $n_1(SU(N)) = N$. These formulas can also be obtained using the classification of holonomy saddles obtained from applying Borel--de Siebenthal theory.

Finally, let $N_2(G)$ be the set of $G$-conjugacy classes of commuting triples $(s,t,u)$ in $G$, such that $s,t$ are semisimple, $u$ is unipotent, and their simultaneous centralizer $G_{s,t,u}$ contains no nontrivial torus.  A result of Jakob--Yun relates $N_2(G)$ to the set of isomorphism classes of indecomposable $G$-bundles on an elliptic curve.  Again, we can compute the number
of bound states in the three-dimensional theory in terms of the number of bound states in two-dimensional theories,
\begin{align*}
N_2(G) & \leftrightarrow \coprod_{[s] \in \Sigma(G)} N_1(G^{\circ}_s) \\
n_2(G) & = \sum_{[s] \in \Sigma(G)} n_1(G^{\circ}_s). \\
\end{align*}
We list the number of threshold bound states for exceptional groups in Table~\ref{tab:boundstates}.
\begin{table}[htp]
\begin{center}
\begin{tabular}{cccc}
\toprule
$G$ & $n_0(\mathfrak{g})$ & $n_1(G)$ & $n_2(G)$ \\
\midrule
$E_6^{\text{sc}}$ & 3 & 13 & 66 \\
\hline
$E_7^{\text{sc}}$ & 6 & 22 & 102 \\
\hline
$E_8^{\text{sc}}$ & 11 & 31 & 113 \\
\hline
$F_4$ & 4 & 10 & 29 \\
\hline
$G_2$ & 2 & 4 & 9 \\
\bottomrule
\end{tabular}
\end{center}
\caption{Threshold bound states in $d$-dimensional SYM on $T^{d-1}$ for $d = 1,2,$ and $3$.}
\label{tab:boundstates}
\end{table}%
Similarly, the bound states in four-dimensional $\cN = 4$ supersymmetric Yang--Mills on $T^3$ receive contributions from $n_2(G)$, but there is a surprise --- $n_3(G)$ also receives contributions from commuting triples
\cite{Witten:1997bs, Kac:1999gw, Witten:2000nv}.

We now consider the bound-state counting problem for the proposed low-energy limit of two-dimensional $\cN = (8,8)$ supersymmetric Yang--Mills theory.  The total number of bound states will be the sum over all orbifold sectors and massive vacua.  The number of ground states in each orbifold sector is easily computed using spectral flow and Chen--Ruan orbifold cohomology \cite{MR2104605} inspired by string theory \cite{Dixon:1985jw}.
The number of ground states in each orbifold $\left( \mathfrak{z}(\mathfrak{l}) \otimes \bC^4 \right)/W_{G,L}$ is $n_{\elliptic}(W_{G,L})$.  The orbifold cohomology of $\bC^4/\bZ_2$ has an interesting implication for bound states \cite{Dijkgraaf:1997vv, Dijkgraaf:1997ku}.  Since the orbifold has no crepant resolution, conformal field theory predicts that it must have a non-commutative resolution.  The general case of non-commutative resolutions has been intensively studied following \cite{MR1792746}.  Applications of these singularities to four-dimensional gauge theories appear in \cite{Cecotti:2025lns}.

Equality of the UV gauge theory and IR orbifold conformal field theory bound state counts implies
\begin{equation*}
n_1(G) = \sum_{(L, \mathcal{E})} n_{\elliptic}(W_{G,L}),
\end{equation*}
which is easily verified \cite{MR1369407}.  For $G = SU(N)$, the identity follows from Gauss's identity
\begin{equation*}
N = \sum_{d | N} \varphi(d),
\end{equation*}
from {\it Disquisitiones Arithmeticae}.
The function $\varphi(d)$ is Euler's totient function, which counts the number of positive integers up to $d$ that are relatively prime to $d$.
The values $n_{\elliptic}(W_{G,L})$ are listed for exceptional groups in Table~\ref{tab:ellipticweyl}.
We list the criterion for the existence of cuspidal pairs in Table~\ref{tab:cuspidal}.  Some of the relevant data for the generalized Springer correspondence is listed in Table~\ref{tab:relativeweyl}.

\begin{table}[htbp]
\centering
\small
\begin{tabular}{lccccc}
\toprule
$G$ & $E_6$ & $E_7$ & $E_8$ & $F_4$ & $G_2$ \\
\midrule
$n_{\elliptic}(W_G)$ & 5 & 12 & 30 & 9 & 3 \\
\bottomrule
\end{tabular}
\caption{Number of elliptic conjugacy classes in the exceptional Weyl groups.}
\label{tab:ellipticweyl}
\end{table}

\begin{table}[htbp]
\centering
\small
\begin{tabular}{llll}
\toprule
$G$ &  \\
\midrule
$\SU(n)$ & $\chi$ is of order $n$\\
$\Spin(2n + 1)$ & $\chi = 1, 2n + 1 \in \square$ \\
& $\chi \neq 1, 2n + 1 \in \triangle$ \\
$\Sp(2n)$ &  $\chi = 1, n \in \triangle$ and $n$ even \\
& $\chi \neq 1, n \in \triangle$ and $n$ odd \\
$\Spin(2n)$ & $\chi = 1, 2n \in \square$ and $n/2$ even \\
& $\chi \neq 1, \chi(\epsilon) = 1, 2n \in \square$ and $n/2$ odd \\
& $\chi(\epsilon) \neq 1, 2n \in \triangle$ \\
$E_6$ & $\chi \neq 1$ \\
$E_7$ & $\chi \neq 1$ \\
$E_8$ & $\chi = 1$ \\
$F_4$ & $\chi = 1$ \\
$G_2$ & $\chi = 1$ \\
\bottomrule
\end{tabular}
\caption{Conditions on $\chi$ for the existence of cuspidal pairs for groups in characteristic $0$
\cite{MR732546, MR803339}.  Here, $\square$ denotes the set of positive square numbers $\{1, 4, 9, 16, \dots \}$, and $\triangle$ denotes the set of positive triangular numbers $\{1, 3, 6, 10, \dots \}$.
$\epsilon$ is in the kernel of the natural map $\Spin(N)\to\SO(N)$; there is one character $\chi$
of $Z(\Spin(N))$ with $\chi(\epsilon)=-1$ if $N$ is odd and 2 if $N$ is even.
Pairs with $\chi(\epsilon)=1$ are the $\SO(N)$ row.  The local system is
rank one except for $\Spin(N)$, where it comes from the even Clifford algebra.}
\label{tab:cuspidal}
\end{table}
\begin{table}[htbp]
\centering
\small
\setlength{\tabcolsep}{4pt}
\begin{tabular}{l>{\raggedright\arraybackslash}p{0.50\textwidth}>{\raggedright\arraybackslash}p{0.33\textwidth}}
\toprule
$G$ & $(W_{G,L},\ \chi)$ & Non-principal unipotent orbits $\cO_L$  \\
\midrule
$\SU(n)$ & $\varphi(d)$ copies of $(S_{n/d},\ \chi\text{ of order }d)$,
  $d\mid n$ & $(k/d,n/d)$ strings\\
$\Sp(2n)$ & $(W(B_{n-T}),\ (-1)^{T})$ for triangular $T\le n$
  & $\Sp(2T)$ on $(2k,\dots,2)$\\
$\Spin(5)$ & $(W(B_2),1)$; $(W(B_1),-1)$ & spin orbit\\
$\Spin(7)$ & $(W(B_3),1)$; $(W(B_1),-1)$ & spin orbit\\
$\Spin(9)$ & $(W(B_4),1)$; $(1,1)$; $(W(B_2),-1)$ &
  $(5,3,1)$; spin orbit\\
$\Spin(8)$ & $(W(D_4),1)$; three $(W(B_2),\chi\neq1)$, one per nontrivial
$\chi$, permuted by triality &
$(3,1)$ and its triality images\\
$E_6$ & $(W(E_6),1)$; $(W(G_2),\omega)$, $(W(G_2),\omega^2)$;
$(1,\omega)$, $(1,\omega^2)$ &
$\SL(3)$ cuspidals on $2A_2$; $(E_6(a_3),\varepsilon\boxtimes\chi)$\\
$E_7$ & $(W(E_7),1)$; $(W(F_4),-1)$; $(1,-1)$ &
$\SL(2)$ cuspidal on $(3A_1)''$; $(E_7(a_5),\varepsilon\boxtimes\chi)$\\
$E_8$ & $(W(E_8),1)$; $(1,1)$ & $(E_8(a_7),\varepsilon)$, $\bC$ \\
$F_4$ & $(W(F_4),1)$; $(1,1)$ & $(F_4(a_3),\varepsilon)$, $\bC$ \\
$G_2$ & $(W(G_2),1)$; $(1,1)$ & $(G_2(a_1),\varepsilon)$, $\bC$ \\
\bottomrule
\end{tabular}
\caption{Generalized Springer correspondence for the
classical groups through rank four and the exceptional groups \cite{MR732546,  MR1021493, MR3694647}.}
\label{tab:relativeweyl}
\end{table}
As an illustration, we consider the bound state counting for gauge group $G = \Spin(7)$.  In the gauge theory, the number of bound states is expressed as a sum over the four ways of deleting a node from the affine Dynkin diagram of $\Spin(7)$:
\begin{align*}
n_1(Spin(7)) & = n_0(\fso(7)) + n_0(\fso(7)) + n_0(\fso(4) \times \fso(3)) + n_0(\fso(6)) \\
& = 1 + 1 + 1 + 1 = 4.
\end{align*}
The low-energy orbifold description has two contributions from $(Spin(7), 1)$ and $(Spin(3), -1)$, so 
\begin{equation*}
n_1(Spin(7)) = n_{\elliptic}(W_{Spin(7)}) +  n_{\elliptic}(W_{Spin(3)}) = 3 + 1 = 4. 
\end{equation*}
In this case, we see the equality of the two expressions.  Gauge groups in the classical series can be treating using generating functions.  They have an interesting interpretation in M-theory, which we turn to next.
\section{M-theory Interpretation}
\label{sec:mtheory}
Threshold bound states in one and two dimensions have elegant interpretations in terms of the BFSS matrix model for M-theory \cite{Banks:1996vh} and matrix strings \cite{Dijkgraaf:1997vv, Dijkgraaf:1997ku}.  The M-theory and type IIB pictures can translate complicated bound state counting problems into simpler combinatorial ones.  Witten's analysis of bound states in supersymmetric quantum mechanics was generalized to the orientifold $S^1 \times \R^9/\mathbb{Z}_2$ \cite{Hanany:1999jy, Lee:2017lfw}.  Duality arguments determine the number of bound states for $Sp_{2n}$ quantum mechanics in terms of the generating function
\begin{equation*}
\sum_{n \ge 0} n_0(Sp_{2n}) t^n = f_{C} = \prod_{n \ge 1} (1 + t^n).
\end{equation*}
Similarly, the bound state generating functions in two and three dimensions,
\begin{align*}
\sum_{n \ge 0} n_1(Sp_{2n}) t^n = f^2_{C} = \prod_{n \ge 1} (1 + t^n)^2, \\
\sum_{n \ge 0} n_2(Sp_{2n}) t^n = f^4_{C} = \prod_{n \ge 1} (1 + t^n)^4, \\
\end{align*}
suggest an orientifold interpretation with $2^d$ fixed points from a $\bZ_2$ action on the $d$-torus $T^d.$
The generating functions for Spin groups are listed in \cite{jakob2024countingabsolutelyindecomposablegbundles}.
For two-dimensional theories, the cuspidal components have an elegant interpretation in terms of the Fermi sea.  The orientifold picture was used to solve the bound-state counting puzzle in four-dimensions \cite{Witten:1997bs}.

\section{Elliptic Genera and Elliptic Springer Theory}
\label{sec:elliptic}
The elliptic genus can provide a precision test of the proposed low-energy vacuum structure of two-dimensional $\cN = (8,8)$ supersymmetric Yang--Mills theory.  In this section, we will describe some of the ingredients that will need to be assembled in order to compute the elliptic genus.  Compared to the gauge theory computation, the low-energy orbifold elliptic genus is remarkably simple.  In many ways, the elliptic genus can be viewed as a direct generalization of the bound-state counting problems in the previous sections.  In addition to the bound-state contribution, the elliptic genus also contains contributions from the continuum.  However, we can recover the number of bound states from the elliptic genus by taking the limit $q \rightarrow 0$ and a limit in the $(+, +, -, -)$ chamber of $\fso(8)$ flavor fugacities.  We expect that the generating functions for the elliptic genera will have interesting automorphic properties \cite{Kac:1999av}.

The expression for the gauge theory elliptic genus \cite{MR3177993,MR3302634} has an intricate structure of contributions labeled by Jeffrey--Kirwan (JK) residues \cite{MR1318878}.  These contributions can be elegantly organized using ideas originating in elliptic Springer theory \cite{MR3383168}.  In the simpler case of pure two-dimensional $\cN = (2,2)$ theories, the residues can be organized by
the Jordan–Chevalley decomposition for $G$-bundles on elliptic curves \cite{MR4524601}.  The groups of residues can be labeled by a nilpotent (equivalently unipotent) orbit of $G$ and a commuting pair of holonomies.  The stabilizers are the
$E$-pseudo-Levi subgroups defined by Fr\u{a}\c{t}il\u{a}, Gunningham, and Li \cite{MR4524601}.  Their classification is given by a twice-iterated form of the Borel–de Siebenthal procedure \cite{friedman1998holomorphic,MR1618343,MR1895253}.  The full $\cN = (8,8)$ theory has a much more complicated sum.  Already, for special unitary groups the sum is over four-dimensional partitions.  These sums over partitions are closely related to the elliptic lift of instanton counting in four-dimensional \cite{Nekrasov:2002qd,Nekrasov:2003rj} and eight-dimensional \cite{Nekrasov:2017cih} gauge theories.

As an example of the additional complexities beyond special unitary gauge groups, we first consider degenerating all the way to supersymmetric matrix integrals over a point.  We will then work our way back up to two dimensions.  The $\cN = 4$ matrix integral partition function can be expressed as a sum over distinguished nilpotent orbits \cite{Eager:2019bxq}.  The residues appearing in the $\cN = 4$ supersymmetric quantum mechanics index are indexed by Heckman--Opdam residual points.  The values of the residues are equal to the formal degrees of unipotent discrete series \cite{MR1748271, demartino2022unramifiedsphericalautomorphicspectrum, demartino2022residuedistributionsiteratedresidues, Eager:2019bxq}.
The elliptic form of the residues are similar to expressions appearing in the theory of double affine Hecke algebras (DAHA) \cite{cherednik2022integralformulasdahainner, lenart2023ellipticclassesperiodichecke}.  The full 2d $\mathcal{N} = (2,2)$ elliptic genus has a simple closed-form expression \cite{MR4312365} since the theory flows to a theory of free twisted chirals in the IR \cite{Aharony:2016jki}.  Even in this simpler case, there is currently no proof of the equality between the UV and IR elliptic genera.  The elliptic genus of the two-dimensional $\mathcal{N} = (8,8)$ theory has four copies of the $E$-pseudo-Levi index sets, similar to how four-dimensional partitions appear in the special unitary case.
Conjecturally, the $q \rightarrow 0$ limit of the elliptic genus can be expressed in terms of the rational invariants $\Omega_{\cN = 16}$ of Yi--Lee \cite{Lee:2016dbm}.  These invariants can in turn be expressed in terms of sums over Binegar's combinatorial Bala--Carter diagrams or by refining Kac--Smilga's analysis \cite{Kac:1999av}.

Compared to the gauge theory computation, the elliptic genus in the IR orbifold description is extremely simple.
The elliptic genus can be evaluated using standard orbifold conformal field theory results; it is given by a sum over pairs of elements in the Weyl group up to simultaneous conjugation.  These sums simplify even further since the expression only depends on the two conjugacy classes to which the pair of elements belong.  While this equality has been computationally verified at low-rank, the equality in general is a physical corollary of the proposed IR equivalence.

These residue computations rhyme with Langlands' contour shift introduced in his work on the functional equations satisfied by Eisenstein series \cite{MR579181}.  For a recent introduction see \cite{MR4943615, hegde2026rethinkingworklanglandseisenstein}.  
Kapustin and Witten related electric-magnetic duality in four-dimensional $\mathcal{N} = 4$ Super Yang--Mills theory to the geometric Langlands program.  Since reducing this theory to two-dimensions results in the two-dimensional $\cN = (8,8)$ theory, it is perhaps unsurprising that this theory is also closely related to the Langlands program.  Lusztig's generalized Springer correspondence also appears in the closely related problem of classifying surface operators in the four-dimensional $\cN = 4$ supersymmetric Yang--Mills theory \cite{Balasubramanian:2014jca}.  Perhaps most surprising is that the two-dimensional theory appears to be related to the Local Langlands Conjectures and \cite{ginzburg1995residueconstructionheckealgebras, MR1413870}.
One hint in this direction is that the number of threshold bound states $n_1(G)$ equals the number of $L$-packets of $G,$
which has deep arithmetic significance.

The implications of the proposed vacuum structure goes beyond the equality of partition functions and bound states.  Recent work has shown the non-renormalization of Wilson coefficients \cite{Bajaj:2026oja}.  We expect that the equivalence gives an isomorphism of holomorphic factorization algebras.  This might organize and illuminate many mysterious results appearing in the Local Langlands program.
\section*{On the Use of AI}
Claude Opus and Fable, GPT-6 Astra, and Gemini 3 were extensively used for suggesting references and typesetting tables.  Claude Opus and Fable verified several of the conjectured relations between bound states using CHEVIE \cite{GH96} and between elliptic genera using SymPy \cite{SymPy} to high orders using highly optimized and parallelized computation over finite fields with minimal guidance.  Refine was used for final proofreading.
\section*{Acknowledgments}
The author would like to thank F. Benini, D. Ben-Zvi, K. Hori, D. Morrison, Y. Tachikawa, and P. Yi for valuable discussions and communications.  The author would especially like to thank Joe Polchinski for fostering his interest in this problem.

\cleardoublepage 
\bibliographystyle{ytphys}
\bibliography{vacuum.bib}
\end{document}